\documentstyle[11pt,aaspp4]{article}

\begin{document}

\title{The Initial Mass Functions in the Super Star Clusters \\
NGC 1569A and NGC 1705-1}

\author{Amiel Sternberg\altaffilmark{1}}
\affil{School of Physics \& Astronomy, Tel Aviv University,
Ramat Aviv, 69978, Israel}
\authoremail{amiel@wise.tau.ac.il}
\altaffiltext{1}{On leave at Department of Astronomy, University of California, Berkeley, CA 94720-3411
amiel@astro.berkeley.edu}

\begin{abstract}

I use recent photometric and stellar velocity dispersion measurements of the
super-star-clusters (SSCs) NGC 1569A and NGC 1705-1 to determine their present-day 
luminosity/mass ($L_V/M$) ratios.  I then use the inferred $L_V/M$ ratios,
together with population synthesis models of evolving star-clusters, to
constrain the initial-mass-functions (IMFs) in these objects.

I find that $(L_V/M)_\odot=28.9$ in 1569A, and $(L_V/M)_\odot=126$ in 1705-1. It follows that in
1569A the IMF is steep with $\alpha\sim 2.5$
for $m^{-\alpha}dm$ IMFs which extend to 0.1 $M_\odot$.
This implies that most of the stellar mass in 1569A is contained in low-mass ($< 1$ $M_\odot$) stars.
However, in 1705-1 the IMF is either flat, with
$\alpha\lesssim 2$, or it is truncated at a lower mass-limit between 1 and 3 $M_\odot$.

I compare the inferred IMFs with the mass functions (MFs) of
Galactic globular clusters.
It appears that 1569A has a sufficient reservoir of low-mass stars for it to plausibly
evolve into an object similar to
Galactic globular clusters. However, the apparent deficiency of low-mass stars in
1705-1 may make it difficult for this SSC to become a globular cluster.
If low-mass stars do dominate the cluster mass in 1705-1, the 
large $L_V/M$ ratio in this SSC may be evidence that the most massive stars have
formed close to the cluster cores.

\end{abstract}


\section{Introduction}

Recent optical and ultraviolet {\it Hubble Space Telescope} ({\it HST})
and near-IR speckle observations 
have revealed the widespread presence of luminous ($L_V\gtrsim10^7$ $L_\odot$) and compact ($r<2$ pc) 
``super-star-clusters" (SSCs) in a variety of star-forming galaxies 
\markcite{Holtzman92} \markcite{Whitmore95} \markcite{Maoz96}
(e.g. Holtzman et al. 1992; Whitmore \& Schweizer 1995; Maoz et al. 1996;
\markcite{Tacconi96} Tacconi-Garman, Sternberg \& Eckart 1996). 
The observations suggest
that in starburst galaxies a large fraction of the OB stars are formed in compact SSCs, 
as opposed to more diffuse stellar associations.

The various photometric and spectroscopic observations are sensitive mainly to the
massive stars ($\gtrsim 3$ $M_\odot$).
Lower mass stars are generally not detectable directly. The {\it total} stellar masses of
the SSCs are usually estimated using models which predict the luminosity/mass
ratios for an assumed set of parameters such as the cluster age and initial mass
function (IMF).  In most models the luminosity is dominated by the most massive (and observed) stars in the
system, while the mass is dominated by the more numerous lower mass (and unobserved) stars.  
Application of such methods has led to inferred cluster masses ranging from 10$^4$ to 10$^6$ $M_\odot$
depending on the particular SSC and model employed.  The large inferred masses and the small
observed radii have led to suggestions that the SSCs are young globular clusters.

NGC 1569 and NGC 1705 are two nearby dwarf galaxies. Each contains a prominent and well studied SSC,
designated 1569A and 1705-1
\markcite{Arp85}\markcite{Melnick85}
(Arp \& Sandage 1985; Melnick, Moles \& Terlevich 1985).
\markcite{Ho1996a}\markcite{Ho1996b} Ho \& Filippenko (1996a,b) carried out high-resolution
(Keck) spectroscopy of 1569A and 1705-1, and were able to measure the stellar velocity dispersions
in these SSCs. The stellar velocities together with the small
cluster sizes indicated by the {\it HST} observations
\markcite{O'Connell94}\markcite{Meurer95}\markcite{DeMarchi97}
(O'Connell, Gallagher \& Hunter 1994; Meurer et al. 1995; DeMarchi et al. 1997)
imply cluster crossing times much shorter than the likely cluster ages.
Ho \& Filippenko concluded that the clusters are gravitationally bound,
and that the implied virial masses are as large as $\sim 10^5$ $M_\odot$.
Ho \& Filippenko also argued that 1569A and 1705-1 might evolve into objects similar to
Galactic globular clusters.

The independent estimates of the total cluster masses made possible by Ho \& Filippenko's
velocity dispersion measurements can be used to ``invert" the usual analysis of
the SSCs.
In this paper I derive
luminosity/mass ratios ($L_V/M$) for 1569A and 1705-1, based on the available observations.  I then compare the
observed $L_V/M$ ratios with model predictions for a wide range of initial conditions.  My main goal
is to constrain the IMFs in these clusters.
I also address the question of whether the inferred IMFs
are consistent with the observed mass-functions (MFs) of present-day
globular clusters.

\section{Models}

Figures 1-4 display ``population synthesis" computations 
of the time-dependent values of
$L_V/M$ for young clusters with ages between 1 and 100 Myr.
In these models it is assumed that all of the stars form in a single
instantaneous ``burst."
Fig. 1 shows the behavior for solar metallicity clusters with power-law
($m^{-\alpha}dm$)
IMFs which extend from a lower-mass limit $m_l=0.1$ $M_\odot$ to an upper-mass
limit $m_u=120$ $M_\odot$.  Results are displayed for 
$\alpha$ ranging from 1.5 to 2.5 
($\alpha=2.35$ for the \markcite{Salpeter55} Salpeter (1955) IMF).
Fig. 2 shows the behavior for low-metallicity ($0.2\times$solar) clusters.
Fig. 3 displays the evolution of $L_V/M$ for a Miller-Scalo IMF
\markcite{Miller79}\markcite{Scalo86} (Miller \& Scalo 1979; Scalo 1986)
and the IMF in the Galactic star-forming region NGC 3603 
(Eisenhauer et al. 1998, see also \S 5),
for both solar and low-metallicity clusters. The small wiggles in the
model curves are numerical artifacts.

Power-law IMFs with $\alpha < 2$ are ``flat" and are biased toward massive stars in the
sense that the cluster mass diverges as $m_u$ becomes large.
IMFs with $\alpha > 2$ are ``steep", and the cluster mass diverges as $m_l$ becomes small.
Table 1 lists the initial mass fractions, $f_M(<1)$, and stellar number fractions,
$f_N(<1)$, contained in stars with masses less than 1 $M_\odot$
for each of the IMFs diplayed in Figs. 1-3. Table 1 also lists the
mean stellar masses, $<m>$, for each of the IMFs.

Figs. 1 and 2 shows that for a given IMF $L_V/M$ 
reaches a maximum value at $\sim 4$ Myr, and then decreases steadily
afterward. 
The peak values of $L_V/M$ are smaller in clusters which form stars sequentially
rather than coevally.
At any time the luminosity is produced by the most massive
cool supergiants and the hot upper part of the main-sequence.  For example,
for a total luminosity $L_V=10^7$ $L_\odot$ at 10 Myr about half of the
V-light is produced by $2\times 10^2$ to 
$4\times 10^2$  K and M supergiants (20-25 $M_\odot$),
and the other half by $5\times 10^3$ to $15\times 10^3$ early B-type stars (5-20 $M_\odot$).
At a fixed age $L_V/M$ is smaller for
steeper IMFs, i.e. for larger values of $\alpha$, due to the larger fractions of low-mass stars.
The luminosity peak occurs when the most massive ($\sim 100$ $M_\odot$)
stars evolve off the main-sequence.  
The luminosity decreases as the massive stars disappear and the ``turn-off"
point moves down the main-sequence. 
In low metallicity clusters the luminosity maxima are larger (by factors of $\sim 2$) 
because a greater number of very massive cool supergiants (rather than
Wolf-Rayet stars) are formed \markcite{Schaerer93}(Schaerer et al. 1993).
The models presented here include mass-loss due to supernova explosions.
Only 1.4 $M_\odot$ stellar remnants are assumed to remain after the
core-collapse explosions.

For steep IMFs the cluster masses are sensitive to
the assumed lower mass limit.  For ages less than 100 Myr, and $m_l$ less than $\sim 3$ $M_\odot$,
the luminosity/mass
ratios are equal to $f$ times the values of $L_V/M$ displayed in Figs. 1 and 2. 
The mass correction factor $f\equiv M(0.1)/M(m_l)$ where $M(m_l)$ is the
cluster-mass assuming the IMF extends to $m_l$ rather than to 0.1 $M_\odot$.
Because of the supernovae, $f$ varies slightly with time.
Fig. 4 displays the correction factors at 10 Myr.

The computations were carried out
using the Geneva stellar evolutionary tracks
\markcite{Schaerer93}(Schaerer et al. 1993), and bolometric corrections
for dwarfs, giants and supergiants compiled by \markcite{Schmidt82}Schmidt-Kaler (1982).  
I will present further details of these and related computations elsewhere
(Sternberg et al. 1998).  Similar models have been presented by others
\markcite{Bruzual93}\markcite{Leitherer95}\markcite{Charlot96}
(e.g. Bruzual \& Charlot 1993 (BC93); Leitherer \& Heckman 1995; Charlot 1996),
and for the same input parameters the various models are in excellent
agreement for ages between 1 and 100 Myr.  A comparison is given in Table 2 which lists computed values of
$(L_V/M)_\odot$ at the peak luminosities, and at 10 Myr, assuming a Salpeter IMF.

\section{Constraining the IMFs in 1569A and 1705-1}

The observed luminosities of 1569A and 1705-1 together with measurements of
their radii and stellar velocity dispersions may be used to determine the luminosity/mass
($L_V/M$) ratios for these sytems.  The $L_V/M$ ratios may then be compared with
the models to constrain the IMFs in these systems. A summary of the cluster properties is listed in Table 3.

I adopt the distances of 2.5 and 5.0 Mpc to NGC 1569 and NGC 1705 which
\markcite{O'Connell94}
O'Connell et al. (1994) derived from color-magnitude fitting of individual stars in 
these galaxies.
Measurements of the nebular oxygen abundances yield
metallicities of 0.2 and 0.45 (relative to solar) for these systems
\markcite{Devost97}\markcite{Kobulnicky97}\markcite{Marloew95}
(Devost, Roy \& Drissen 1997; Kobulnicky \& Skillman 1997; Marlowe et al. 1995).

Recent optical {\it HST} imaging of 1569A by \markcite{DeMarchi97} 
De Marchi et al. (1997) shows that it actually consists of two 
components with an angular separation of only 0.2$\arcsec$.
\markcite{Gonzalez97}
Gonz\'{a}lez-Delgado et al. (1997) detected HeII and NIII Wolf-Rayet (WR)
emission features in addition to the Ca II red-supergiant (RSG) photospheric
absorption triplet.
De Marchi et al. suggested that the WR stars and RSGs each trace a separate component.
The presence of WR stars implies a cluster age of $\sim 4$ Myr,
while the presence of RSGs implies an age of
$\gtrsim 10$ Myr.  
I assume here that the RSG (one-dimensional) velocity dispersion
$\sigma=15.7$ km s$^{-1}$ that \markcite{Ho96a} Ho \& Filippenko (1996a)
observed in 1569A is of the
brighter component which (for a distance of 2.5 Mpc) has a luminosity
$L_V=3.1\times10^7$ and a half-light radius $r_h=1.8$ pc
\markcite{DeMarchi97}
(De Marchi et al. 1997). The crossing time, $t_c\equiv2r_h/\sigma=2.3\times 10^5$ yr
is much shorter than the cluster age. Therefore, the SSC is likely gravitationally bound.
For a bound system the virial
mass, $M\equiv 3\sigma^2R/G$ where $R$ is the gravitational radius, and
for a wide range of virialized stellar mass distributions $M\approx 10\sigma^2r_p/G$
where $r_p$ is the projected half-mass radius \markcite{Spitzer87}(Spitzer 1987).
Assuming that light traces mass, so that $r_p=r_h$, I infer a virial mass of $1.1\times 10^6$ $M_\odot$,
giving $(L_V/M)_\odot=28.9$ for 1569-A. This value of $L_V/M$ is indicated in Figs. 1-3.

The more distant cluster 1705-1 may be a simpler system 
\markcite{Meurer92} (Meurer et al. 1992).  Ultraviolet {\it HST} 
spectroscopy reveals an absence of WR and O-type stars,
suggesting that 1705-1 is a $\sim 10$ to 20 Myr old ``postburst" object 
\markcite{Heckman97}(Heckman \& Leitherer 1997). 
\markcite{O'Connell94}O'Connell et al. (1994)
measured an optical half-light radius of 3.4 pc and a cluster luminosity $L_V=3.4\times 10^7$ $L_\odot$.
\markcite{Meurer95}Meurer et al. (1995) reanalyzed the \markcite{O'Connell94}O'Connell et al. 
data and argued that the half-light radius is actually only 0.9 pc. 
\markcite{Ho1996b}Ho \& Filippenko (1996b) measured an RSG velocity
dispersion of 11.4 km s$^{-1}$ giving a crossing time of $1.6\times 10^5$ yr. For
$r_p=0.9$ pc the virial mass is $2.7\times 10^5$ $M_\odot$,
so that $(L_V/M)_\odot=126$ in 1705-1.  This value for $L_V/M$ is also indicated in Figs. 1-3.
The virial mass
is much smaller than the $\sim 10^6$ $M_\odot$ which
Meurer et al. (1995) inferred for 1705-1 using their measurement of the UV-luminosity,
and a model UV-luminosity/mass ratio for a cluster with a 
Salpeter IMF extending to 0.1 $M_\odot$.

I note that Ho \& Filippenko (1996a,b) assumed that the half-light radii
are equal to the gravitational radii (i.e. that $r_h=R$) and derived masses that
are 10/3 times smaller than the virial masses I have inferred.  Their mass estimates
may be regarded as lower limits, and would give $(L_V/M)_\odot=96.3$ for 1569-A, and
$(L_V/M)_\odot=419$ for 1705-1.

The cluster IMFs can now be constrained by comparing the observed $L_V/M$ ratios
with the models.

Figs. 1 and 2 show that 
if 1569A is emitting at close to the peak luminosity then the IMF must be very steep
with $\alpha>2.5$ even if $m_l=0.1$ $M_\odot$.  Second, at the likely cluster age of 10 Myr the
observed $L_V/M$ ratio is consistent with a Salpeter IMF ($\alpha=2.35$) extending to $m_l=0.1$ M$_\odot$.
Third, for cluster ages up to $\sim 20$ Myr, the predicted $L_V/M$ ratios for a Miller-Scalo IMF
are larger than the observed values. These three conclusions imply that a large fraction
of the stellar mass in 1569A is contained in low-mass stars. For example, if $\alpha=2.35$
then 60\% of cluster mass is contained in stars with masses less than 1 $M_\odot$ (see Table 1).

On the other hand, 1705-1 appears to be deficient in low-mass stars. First, the IMF in this cluster
is consistent with a Salpeter IMF with $m_l=0.1$ $M_\odot$ only if it is emitting
at the peak luminosity.  However, this is unlikely since many ($\sim 5\times 10^3$) O and WR stars would then be present.
Second, for a Miller-Scalo IMF the observed $L_V/M$ ratio in 1705-1 is larger than the predicted values
at all cluster ages.  Third, for a likely cluster age between 10 and 20 Myr, the measured $L_V/M$ ratio
is consistent with a Salpeter IMF only if the IMF is 
truncated at values of $m_l$ ranging from $\sim 1$ to 3 $M_\odot$ (see Fig. 4).
Alternatively, for ages between 10 and 20 Myr, the $L_V/M$ ratio is consistent with 
flat IMFs with $\alpha$ between 2 and 1.5.
For these values of
$\alpha$ the initial cluster mass fraction in $m<1$ $M_\odot$ stars ranges from 32\% to 6\%
(see Table 1). 
It appears that the IMF in 1705-1 is biased towards high-mass stars.

\section{Comparison with Globular Clusters}

The large masses and small radii of the luminous SSCs,
and the fact that they appear to be gravitationally bound objects, has led to
suggestions that they may be young globular clusters
\markcite{Lutz91}\markcite{Holtzman92}\markcite{Larson88}
(Larson 1988; Lutz 1991; Holtzman et al. 1992).  However, this idea has remained controversial
\markcite{Meurer95b}\markcite{vandenBergh95}
(Meurer 1995; van den Bergh 1995).

Ho \& Filippenko (1996a,b) applied the BC93 models 
to 1569A and 1705-1 and concluded that after fading for 10-15 Gyr these objects would
attain $(L_V/M)_\odot$ ratios close to the values of $\sim 0.5-1$ observed in Galactic globular clusters
\markcite{Mandushev91} (Mandushev, Spassova, \& Staneva 1991).
However, Ho \& Filippenko assumed that the cluster masses are
significantly smaller than their likely virial masses (see \S 3.). They also assumed that the
SSCs are presently emitting at their peak luminosities.
Furthermore, the BC93 models are restricted to a Salpeter IMF (with $m_l=0.1$ $M_\odot$), and
predict $(L_V/M)_\odot \sim 0.1$ for 10-15 Gyr old clusters.  
Finally, the BC93
models do not account for cluster mass-loss due to either stellar evolution
or dynamical processes. 

An alternative approach is to ask whether
Galactic globular clusters could have evolved from objects with IMFs which are
constrained by the $L_V/M$ ratios observed in 1569A and 1705-1.

 
Recent {\it HST} observations of Galactic globular clusters probe the luminosity
functions and MFs from the turn-off mass ($\sim 0.8$ $M_\odot$) down nearly to the hydrogen burning limit
\markcite{King98}
(King et al. 1998). The observations indicate that 
for masses in the range $0.1 \lesssim m \lesssim 0.5$ $M_\odot$ the MFs may
be represented as power-laws $m^{-\alpha}dm$ with $\alpha$ between $\sim 0.5$ and 1
(Piotto, priv. com. 1998). 
Luminosity profile studies and dynamical modelling \markcite{Meylan91} (e.g. 
Meylan \& Mayor 1991), as well as pulsar studies
\markcite{Kulkarni90}(Kulkarni, Narayan \& Romani 1990)
provide limits on the globular cluster mass-fractions contained in
more massive evolved stellar components such as white dwarfs and neutron stars. 

To compare globular cluster MFs with the IMFs in 1569A and 1705-1, I list in 
Table 4 the $L_V/M$ ratios for 10 Myr old clusters assuming IMFs (labelled ``A") which vary as $m^{-0.7}dm$
for stellar masses $0.1\le m<0.8$ $M_\odot$, and as $m^{-\alpha}dm$ for $0.8\le m < 120$ $M_\odot$,
as well as IMFs (labelled ``B") which vary as $m^{-\alpha}dm$ for the entire range of 0.1 to 120 $M_\odot$.  
The ``A" type IMFs are defined so that the distribution of $m<0.8$ $M_\odot$ stars initially
resemble the observed (low-mass) MFs of Galactic globular clusters, with additional power-law components
extending to the highest mass stars.  Clusters with ``B" type IMFs initially
contain much more mass in $m<0.8$ M$_\odot$ stars than do present-day globular clusters.

After $\sim 15$ Gyr the cluster masses will decrease as the
massive stars evolve and lose mass.
Table 4 lists the mass ratios $M_{GC}/M$, where $M$ is the SSC mass at 10 Myr,
and $M_{GC}$ is the ``globular cluster" mass at 15 Gyr assuming 
that all stars with initial masses $m>8$ $M_\odot$ become 1.4 $M_\odot$ neutron stars, and that stars
with initial masses in the ranges 0.8-1.5, 1.5-2.5, and 2.5-8 $M_\odot$ become white dwarfs with
masses equal to 0.6, 0.7, and 1.1 $M_\odot$ respectively.
Table 3 also lists the mass fractions $f(<0.8)$, $f(WD)$ and $f(neutron)$, of the mass
$M_{GC}$ contained in $m<0.8$ $M_\odot$ main-sequence stars, white dwarfs, and neutron stars.  
In their dynamical study \markcite{Meylan91}Meylan \& Mayor (1991) concluded that 
$f(<0.8)\sim 0.73$, $f(WD) \sim 0.25$ and $f(neutron) \sim 0.02$. The dynamical estimate for $f(neutron)$ is
consistent with the number of neutron stars per unit globular cluster mass inferred from
millisecond pulsar observations \markcite{Kulkarni90}(Kulkarni et al. 1990). 
Table 4 can be used to select an IMF which is consistent with the observed $L_V/M$
ratios in 1569A and 1705-1, and to then determine how these SSCs must evolve if
they are to become objects with MFs similar to those in Galactic globular clusters.

The $L_V/M$ ratio of 28.9 in 1569A is too small to be compatible with any of the ``A" type IMFs
(assuming a cluster age of $\sim 10$ Myr).
As discussed in \S 3 the small observed luminosity/mass ratio implies a large mass fraction
in low mass stars, such as a ``B" type IMF with $\alpha=2.5$.  For this IMF, evolution
of the massive ($m>0.8$ $M_\odot$) stars will reduce 
1569A to 0.81 of its present mass (see Table 4). In addition, 1569A must lose
at least 80\% of the mass it now contains in low-mass stars for the MF to flatten to
$m^{-0.7}dm$ between 0.1 and 0.8 $M_\odot$. About half of the mass contained in WD and neutron star
remnants would also have to be removed for the final MF to be consistent with the distribution
$f(<0.8)=0.73$, $f(WD)=0.25$ and $f(neutron)=0.02$.  Dynamical studies show that processes such as
tidal stripping and cluster evaporation are effective at removing low-mass stars from evolving globular clusters
\markcite{Spitzer87}\markcite{Chernoff90}(Spitzer 1987; Chernoff \& Weinberg 1990),
although the tidal forces acting on 1569A may be quite different from those acting on
Galactic globular clusters. In any event, it appears that mass loss due to stellar evolution and dynamical
processes must reduce 1569A to $\sim 0.81\times0.24=0.19$ of its present mass if it is
to evolve into an object similar to Galactic globular clusters.
The final cluster mass would be $2.0\times 10^5$ $M_\odot$.

As discussed in \S 3 the $L_V/M$ ratio of 126 in 1705-1 implies that if the IMF
is steep ($\alpha>2$) it must be truncated
at lower mass limits ranging from $\sim 1$ to 3 $M_\odot$. After 15 Gyr 1705-1 would then consist
only of massive stellar remnants.  Alternatively, the IMF in 1705-1 could be flat
with a small mass fraction in low-mass stars. For example, 1705-1 could have an ``A" type IMF
with $\alpha$ between 2.0 and 1.5 (see Table 4).  If $\alpha=2$, evolution of the massive
stars will reduce 1705-1 to 0.43 of its present mass. About 70\% of the WDs and 90\% of
the neutron stars would have to be ejected via dynamical processes for the final MF
to be consistent with $f(<0.8)=0.73$, $f(WD)=0.25$ and $f(neutron)=0.02$.   
The total mass loss would then reduce 1705-1 to $0.43\times 0.53=0.25$ of its present mass, and the
final cluster mass would be $6.8\times 10^4$ $M_\odot$. 
If $\alpha=1.5$ the cluster mass would decrease to $0.12$ of the present mass, for a final mass
of $3.2\times 10^4$ $M_\odot$.  In this scenario
all of the low-mass stars would have to be retained during the cluster evolution, since the MF is of the form
$m^{-0.7}dm$ initially. High velocity kicks imparted by the core collapse explosions could remove the neutron stars 
(\markcite{Drukier96}\markcite{Fryer98}Drukier 1996; Fryer, Burrows \& Benz 1998).
Removal of the WDs will be more difficult since tidal stripping and cluster evaporation
preferentially remove the {\it low-mass} stars
\markcite{Chernoff91}(Chernoff \& Weinberg 1991). It appears that 1705-1 may not 
evolve into a globular cluster. If it does, dynamical mass losses must lead
mainly to the removal of the massive stellar remnants.

\section{Discussion}

I have used the observed values of $L_V/M$ in 1569A and 1705-1 to constrain their IMFs.
In 1569A the IMF must be steep, and a large fraction of the SSC mass is contained in
low-mass ($m<1$ $M_\odot$) stars. This cluster could plausibly evolve into a
$2\times 10^5$ $M_\odot$ object resembling Galactic
globular clusters. However, in 1705-1 the IMF must be flat or truncated at a lower-mass limit
exceeding 1 $M_\odot$. 
Most of the cluster mass in 1705-1 is contained in high-mass stars, implying that
it may be difficult for this SSC to evolve into a globular cluster. 

Several uncertainties could affect my analysis and conclusions.
First, because $L_V/M$ is proportional to
distance, and the distances to 1569A and 1705-1 are uncertain to within factors of $\sim 1.5$
\markcite{Meurer92}\markcite{O'Connell94}(Meurer et al. 1992; O'Connell et al. 1994),
the luminosity/mass ratios could be $1.5$ times smaller or larger than assumed here.
Second, the observed luminosities have been corrected only for Galactic foreground
extinction, and if there is any internal extinction the $L_V/M$ ratios could be much larger.
The observations suggest that the internal extinction is very small
\markcite{Meurer92}\markcite{Gonzalez97}(Meurer et al. 1992; Gonzalez-Delgado et al. 1997).
However, I note that the initial extinction in these objects was probably very large.
The fact that the SSCs are gravitationally bound implies that they must have formed
with high ($\sim 50\%$) gas-to-star conversion efficiencies 
\markcite{Mathieu84}\markcite{Lada84}(Mathieu 1984; Lada et al. 1984).
If the protostellar molecular cloud masses and sizes were approximately
equal to the observed SSC masses and sizes, the initial molecular hydrogen ($H_2$) gas densities
must have been as large as $\sim 10^5$ cm$^{-3}$, with corresponding $H_2$ column densities
equal to $\sim 5\times 10^{23}$ cm$^{-2}$.  For Galactic gas-to-dust ratios these column densities
correspond to visual extinctions $A_V\sim 100$.  So if the internal extinction is now very small
any remaining gas must have been blown out very efficiently once the stars
started forming.  This also implies that the star formation was ``instantaneous" rather than
``continuous" in these SSCs.

Third, the velocity dispersions measured by \markcite{Ho96a}\markcite{Ho96b}Ho \& Filippenko (1996a,b)
were based on observations of massive red supergiants (RSGs).  If these stars have attained equipartition,
and are dynamically segregated in the cluster cores, the RSG velocities would be smaller than the mean stellar
velocities, and the RSG virial masses would underestimate the total cluster masses.
However, equipartition occurs on a cluster relaxation time $t_r\approx (N/8{\rm ln}N)t_c$, where
$N$ is the number of stars in the cluster, and $t_c$ is the crossing time 
\markcite{Bonnell98}(Bonnell \& Davies 1998).  Adopting a mean stellar mass $<m>=1$ $M_\odot$
(see Table 1), and the cluster masses and crossing times derived in \S 3, it follows that
$t_r \gtrsim 100$ Myr.  It appears that the SSCs are too young for the RSGs to have undergone dynamical mass segregation.
Alternatively, the massive stars may have formed preferentially
closer to the proto-cluster cores \markcite{Hillenbrand97}({\it cf.} Hillenbrand 1997).
The velocities could be independent of the stellar masses, but the observed half-light radii would provide
lower limits to the true cluster sizes. The cluster masses would again be underestimated. 
If low-mass ($m<1$ $M_\odot$) stars do dominate the total cluster mass in 1705-1
its large value of $L_V/M$ can be taken as evidence of mass segregation.
I note that if the massive stars are spatially segregated the SSCs should appear more compact in the
UV and near-IR since such light is produced by the most massive stars,
whereas a significant fraction of the optical
light is produced by intermediate mass stars (see \S 2).

Finally, if significant populations of RSGs exist in both components of 1569A (see \S 3) it is possible
that part of the signal measured by Ho \& Filippenko (1996a) is due to relative motion
between the components rather than an intrinsic stellar velocity dispersion. 
The inferred $L_V/M$ ratio for 1569A would then be larger. Additional high-resolution
spectroscopy is required to clarify this point.

Recent stellar census studies of nearby young clusters provide
direct probes of the high-mass portions of the IMFs in star forming regions.  In Galactic and Magellanic OB
associations \markcite{Massey95}Massey et al. (1995) found that $\alpha=2.3\pm 0.3$
between 3 and 120 $M_\odot$. 
Near-IR adaptive optics imaging \markcite{Brandl96}(Brandl et al. 1996) and 
optical {\it HST} imaging \markcite{Massey98}(Massey \& Hunter 1998) of R136,
the ``core" ($r<2$ pc) of the 30 Doradus star-forming region in the LMC, imply
an IMF with $\alpha\sim 2.3$ from 120 $M_\odot$ down to the confusion limit
of about 2 $M_\odot$, with no hint of a break near the observed
low-mass limit.   In the Galactic cluster NGC 3603 
\markcite{Eisenhauer98} Eisenhauer et al. (1998) found that within the central parsec the MF is
actually biased toward massive stars, with $\alpha\sim 1.7$ from the
detection limit of $\sim 1$ $M_\odot$ to $\sim 30$ $M_\odot$, steepening to $\alpha\sim 2.7$
for higher masses. (This steepening may be due to the small cluster mass rather than to
an intrinsic steepening of the IMF).
\markcite{Hillenbrand97}Hillenbrand (1997) found that the Orion Nebula Cluster
is globally consistent with a Miller-Scalo IMF, but that 
within the central 0.3 pc the IMF is biased towards massive stars.
It is unclear what these various observations imply about
the IMFs in the much more massive systems 1569A and 1705-1. 
Nevertheless, Figs. 1-3 illustrate that the IMF in 1569A may be
similar to the global MF in the Orion cluster, whereas the IMF in 1705-1
may be similar to the MF in NGC 3603.

Recent observations of many starburst galaxies indicate that a large, 
and perhaps dominant, fraction of the recent star formation has occured in massive SSCs
\markcite{O'Connell95}\markcite{Maoz96}\markcite{Tacconi96}
(O'Connell et al. 1995; Maoz et al. 1996; Tacconi-Garman et al. 1996).
If most of the SSCs have steep IMFs similar to 1569A, this would imply that dynamically
significant populations of
low-mass stars are formed togther with the observed high-mass stars.
However, if most of the SSCs are similar to 1705-1, with flat or truncated IMFs, this
would support the 
long-standing hypothesis \markcite{Rieke80}\markcite{Rieke93} (Rieke et al. 1980, 1993)
that in starburst galaxies massive stars are formed preferentially.
Additional dynamical measurements of SSC masses would be valuable.

\acknowledgments

I thank Leo Blitz, Alex Filippenko, James Graham, Luis Ho, Ivan King, Dan Maoz, and Chris McKee for discussions,
and the referee for helpful comments.
I thank the Radio Astronomy Laboratory at U.C. Berkeley, the Center for Star Formation Studies consortium,
and the U.S.-Israel Binational Science Foundation (grant 94-00300) for support.






\clearpage
 
\begin{deluxetable}{llll}
\footnotesize
\tablecaption{Initial Mass Functions\label{tbl-1}} 
\tablehead{
\colhead{IMF\tablenotemark{a}} & \colhead{$f_M(<1)$\tablenotemark{b}} &
 \colhead{$f_N(<1)$\tablenotemark{c}} & \colhead{$<m>$\tablenotemark{d}} \nl
}
\startdata
$\alpha=1.5$ & 0.06 & 0.70 & 3.5 \nl
\ \ \ \ \ \ 2.0 & 0.32 & 0.90 & 0.71 \nl
\ \ \ \ \ \  2.35 & 0.60 & 0.96 & 0.35 \nl
\ \ \ \ \ \ 2.5 & 0.70 & 0.97 & 0.29 \nl
Miller-Scalo & 0.31 & 0.70 & 0.79 \nl
NGC 3603 & 0.18 & 0.82 & 1.31 \nl
\enddata

\tablenotetext{a}{Initial mass functions in the range 0.1 to 120 $M_\odot$.}
\tablenotetext{b}{Initial cluster mass fractions in stars less massive than 1 $M_\odot$.}
\tablenotetext{c}{Initial stellar number fractions in stars less massive than 1 $M_\odot$.}
\tablenotetext{d}{Mean stellar masses.}

\end{deluxetable}

\clearpage

\begin{deluxetable}{lllll}
\footnotesize
\tablecaption{Model Comparisons\label{tbl-2}}
\tablehead{
\colhead{} & \colhead{peak} & \colhead{peak} & \colhead{10 Myr} & \colhead{10 Myr} 
\nl
\colhead{} & \colhead{$(L_V/M)_\odot$} & \colhead{$(L_V/M)_\odot$} & 
\colhead{$(L_V/M)_\odot$} & \colhead{$(L_V/M)_\odot$} 
\nl
\colhead{Model\tablenotemark{a}} & \colhead{solar}   & \colhead{0.2$\times$solar} &
\colhead{solar}   & \colhead{0.2$\times$solar}
}
\startdata
this paper & 63 & 117 & 31 & 38 \nl
C96\tablenotemark{b} & 63 & 80 & 27 & 27 \nl
LH95\tablenotemark{c} & 68 & 108 & 35 & 50 \nl
\enddata


\tablenotetext{a}{Salpeter IMF between 0.1 and 120 $M_\odot$.  The C96 and LH95 values have been
adjusted for supernova mass-loss.} 
\tablenotetext{b}{Charlot 1996}
\tablenotetext{c}{Leitherer and Heckman 1995}

\end{deluxetable}

\clearpage
 
\begin{deluxetable}{lllllllll}
\footnotesize
\tablecaption{Super-star-cluster properties \label{tbl-3}}
\tablehead{
\colhead{} & \colhead{D\tablenotemark{a}} & \colhead{Z\tablenotemark{b}} & 
\colhead{$R_h$\tablenotemark{a,c}} & \colhead{$\sigma$\tablenotemark{d}} &
\colhead{$L_V$\tablenotemark{a,c}} & \colhead{$M$} & \colhead{$(L_V/M)_\odot$} &
\colhead{age\tablenotemark{e}}
\nl
\colhead{object} &  \colhead{(Mpc)} & \colhead{(solar)} & \colhead{(pc)} & \colhead{(km s$^{-1}$)} &
\colhead{($L_\odot$)} & \colhead{($M_\odot$)} & \colhead{} & \colhead{Myr}
}
\startdata
1569A  & 2.5 & 0.2 & 1.8 & 15.7 & $3.1\times 10^7$ & $1.1\times 10^6$ & 28.9 & 4-10\nl
1705-1 & 5.0 & 0.45 & 0.9 & 11.4 & $3.4\times 10^7$ & $2.7\times 10^5$ & 126 & 10-20 \nl
\enddata

 
\tablenotetext{a}{O'Connell et al. 1994}
\tablenotetext{b}{Devost et al. 1997, Kobulnicky and Skillman 1997}
\tablenotetext{c}{Meurer et al. 1995, De Marchi et al. 1997}
\tablenotetext{d}{Ho and Fillipenko 1996a,b}
\tablenotetext{e}{Gonz\'{a}lez-Delgado et al. 1997, Leitherer and Heckman 1997}

\end{deluxetable}

\clearpage
\begin{deluxetable}{llllllll}
\footnotesize
\tablecaption{Super-Star-Cluster IMFs vs. Globular Cluster MFs  \label{tbl-4}}
\tablehead{
\colhead{IMF\tablenotemark{a}} & \colhead{$\alpha$} &
 \colhead{$(L_V/M)_\odot$\tablenotemark{b}} & \colhead{$(L_V/M)_\odot$} &
\colhead{$M_{GC}/M$} &  \colhead{$f(<0.8)$} &
\colhead{$f(WD)$} & \colhead{$f(neutron)$}
\nl
\colhead{} & \colhead{} & \colhead{solar} & \colhead{$0.2\times$solar} & \colhead{} &
\colhead{} & \colhead{} & \colhead{}
}
\startdata
A & 1.5 & 187.0 & 287.9 & 0.33 & 0.25 & 0.47 & 0.28 \nl
A & 2.0 & 100.0 & 140.7 & 0.43 & 0.42 & 0.48 & 0.10 \nl
A & 2.35 & 58.3 & 70.7 & 0.53 & 0.52 & 0.44 & 0.04 \nl
A & 2.5 & 40.4 & 59.0 & 0.56 & 0.55 & 0.42 & 0.03 \nl
  &     &      &      &      &      &      &      \nl
B & 1.5 & 175.6 & 270.3 & 0.37 & 0.37 & 0.39 & 0.24 \nl
B & 2.0 & 74.3  & 104.5 & 0.58 & 0.68 & 0.27 & 0.05 \nl
B & 2.35 & 30.9 & 37.5 & 0.75  & 0.82 & 0.17 & 0.02 \nl
B & 2.5 & 17.8  & 26.0 & 0.81  & 0.86 & 0.13 & 0.01 \nl
\enddata

\tablenotetext{a}{IMFs labelled ``A" vary as $m^{-0.7}$ for $0.1\le m < 0.8$ and $m^{-\alpha}$ 
for $0.8\le m < 120$ $M_\odot$.
IMFs labelled ``B" vary as $m^{-\alpha}$ for the entire range $0.1 \le m < 120$ $M_\odot$.}
\tablenotetext{b}{Luminosity/mass ratios at 10 Myr.}

\end{deluxetable}

\clearpage

\begin{figure}[htpb]
\plotone{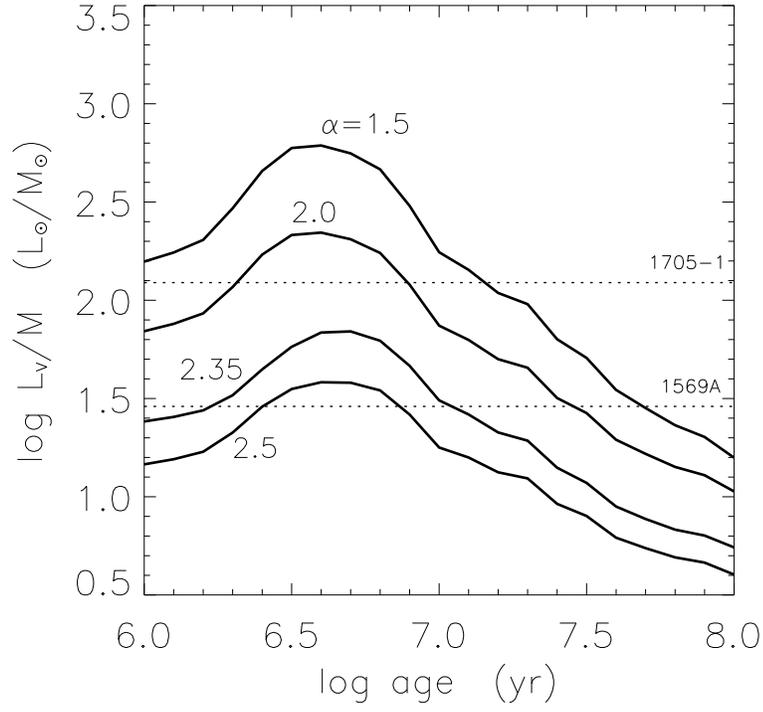}
\caption{$L_V/M$ for solar-metallicity clusters, with power-law ($m^{-\alpha}dm$) IMFs from
0.1 to 120 $M_\odot$. The dotted lines indicate the observed values of $L_V/M$ in 1569A and 1705-1.}
\end{figure}
\clearpage
\begin{figure}[htpb]
\plotone{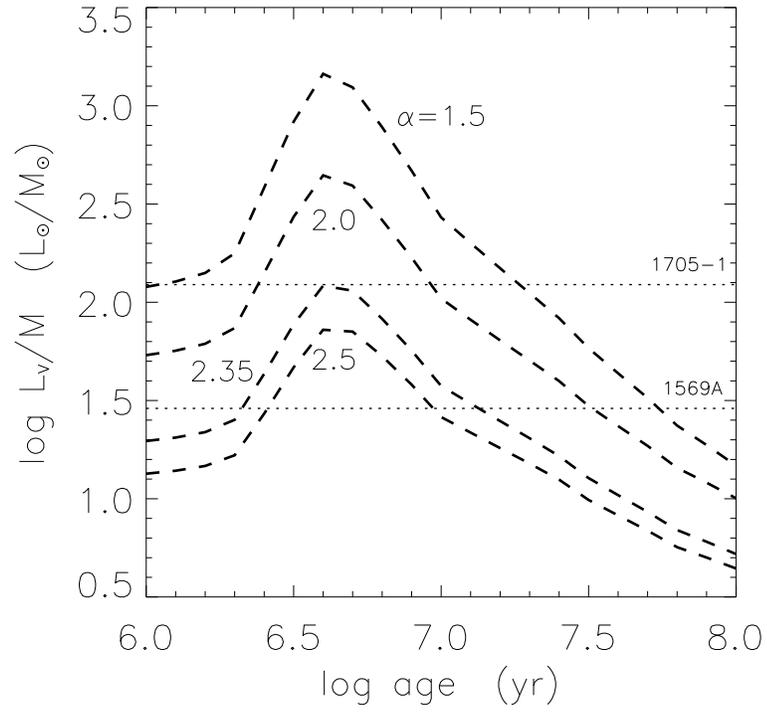}
\caption{$L_V/M$ for low-metallicity (0.2$\times$solar) clusters for power-law IMFs.}
\end{figure}
\clearpage
\begin{figure}[htpb]
\plotone{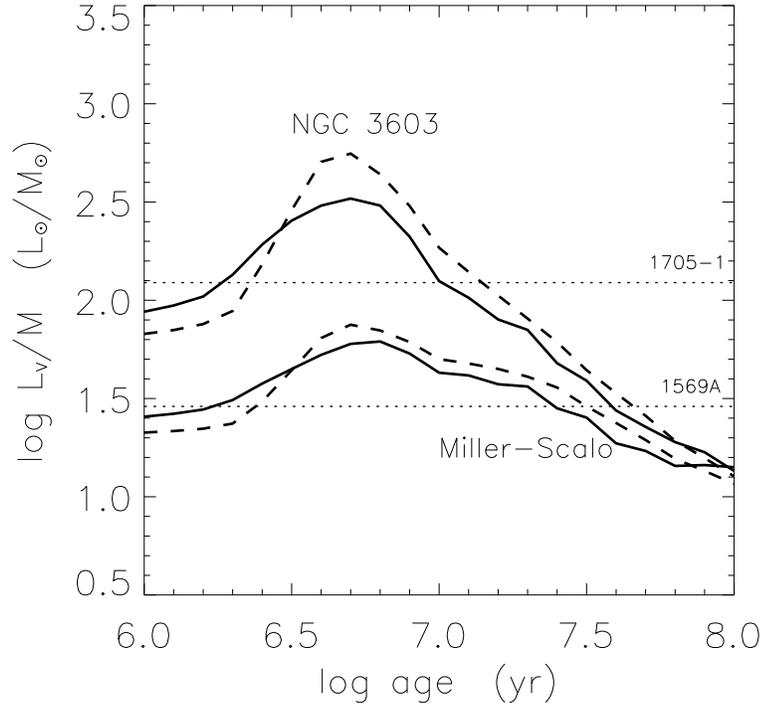}
\caption{$L_V/M$ assuming a Miller-Scalo IMF, and an IMF equal to the
observed MF in NGC 3603 (see text). Solid curves are solar, and dashed curves
are 0.2$\times$solar metallicity.}
\end{figure}
\clearpage
\begin{figure}[htpb]
\plotone{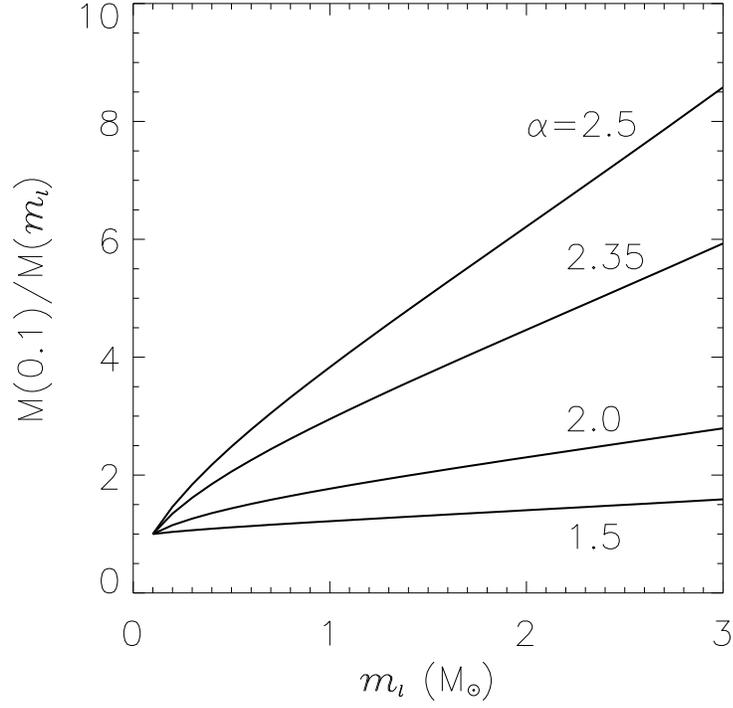}
\caption{Mass correction factor $f\equiv M(0.1)/M(m_l)$ for power-law IMFs at 10 Myr,
where $M(m_l)$ is the cluster mass assuming the IMF extends to a lower mass limit $m_l$,
and $M(0.1)$ is the cluster mass assuming the IMF extends to 0.1 $M_\odot$.}
\end{figure}

\end{document}